\documentclass[american]{article}
\usepackage[T1]{fontenc}
\usepackage[utf8]{inputenc}
\usepackage{float}
\usepackage{units}
\usepackage{amsmath}
\usepackage{graphicx}

\makeatletter

\providecommand{\tabularnewline}{\\}

\usepackage{url}
\usepackage{hyperref}

\makeatother

\usepackage{babel}
\begin{document}
\title{Selected decays of $B_{s}$ meson in covariant confined quark model}
\author{Stanislav Dubni\v{c}ka$^{1}$\\
Anna Zuzana Dubni\v{c}kov\'{a}$^{2}$\\
Mikhail A. Ivanov$^{3}$\\
Andrej Liptaj$^{1\ddagger}$}
\maketitle
\begin{description}
\item [{$^{1}$}] Institute of Physics,\\
Slovak Academy of Sciences, Bratislava, Slovakia.
\item [{$^{2}$}] Faculty of Mathematics, Physics and Informatics,\\
Comenius University, Bratislava, Slovakia.
\item [{$^{3}$}] Bogoliubov Laboratory of Theoretical Physics,\\
Joint Institute for Nuclear Research, Dubna, Russia.
\item [{$^{\ddagger}$}] andrej.liptaj@savba.sk
\end{description}
\begin{abstract}
We present a study of various $B_{s}$ meson decays, including hadronic
and semileptonic final states with different spins and diagram topologies.
The covariant confined quark model is employed to describe hadronic
effects, and our analysis serves as a broad test of the current understanding
of the underlying dynamics within the Standard Model. The level of
agreement with experimental data varies across channels: it is good
for the semileptonic decays $B_{s}^{0}\to D_{s}^{-}\mu^{+}\nu_{\mu}$
and $B_{s}^{0}\to K^{-}\mu^{+}\nu_{\mu}$; acceptable for the hadronic
modes $B_{s}^{0}\to D_{s}^{-}D^{+}$ and $B_{s}^{0}\to K^{-}\pi^{+}$;
marginal for $B_{s}^{0}\to D_{s}^{-}\pi^{+}$, $B_{s}^{0}\to D_{s}^{-}\rho^{+}$,
and $B_{s}^{0}\to\phi\,J/\psi$; and significantly discrepant for
$B_{s}^{0}\to\phi\,\overline{D^{0}}$ and $B_{s}^{0}\to\phi\,\eta_{c}$.
We argue that the observed inconsistencies may arise from the breakdown
of naive factorization and unaccounted long‐distance effects. In particular,
the channels $B_{s}^{0}\to D_{s}^{-}\pi^{+}$ and $B_{s}^{0}\to D_{s}^{-}\rho^{+}$
are theoretically among the cleanest nonleptonic decays, yet they
exhibit persistent discrepancies also reported by other theoretical
groups.
\end{abstract}

\section{Introduction}

Decays of $B_{s}$ mesons are very rich and include a variety of different
spin, flavor and diagram structures. They are usually addressed in
some specific context, our choice to treat them as a whole is mostly
data driven. We use the framework of the covariant confined quark
model (CCQM) \cite{Branz:2009cd} to calculate matrix elements and
observables and we have already applied this approach to the $B_{s}$
particle several times \cite{Ivanov:2011aa,Dubnicka:2013vm,Issadykov:2015iba,Dubnicka:2016nyy,Dubnicka:2018gqg,Issadykov:2018sgw}.
Nevertheless, since our last text, a quantity of new data on $B_{s}$
decays appeared and we believe it is appropriate to come back to the
subject. These data have different nature and include both, hadronic
and semileptonic decays. Specifically, we study the processes
\begin{equation}
\begin{array}{lll}
B_{s}\to D_{s}^{-}\pi^{+}, & \quad B_{s}\to D_{s}^{-}\rho^{+}, & \quad B_{s}\to D_{s}^{-}D^{+},\\
B_{s}\to K^{-}\pi^{+}, & \quad B_{s}\to\phi\bar{D^{0}}, & \quad B_{s}\to\phi\eta_{c},\\
B_{s}\to D_{s}^{-}\mu^{+}\nu_{\mu}, & \quad B_{s}\to K^{-}\mu^{+}\nu_{\mu}, & \quad B_{s}\to\phi e^{+}e^{-},
\end{array}\label{eq:Processes}
\end{equation}
all having established branching fractions \cite{ParticleDataGroup:2024cfk}
as measured by \cite{LHCb:2021qbv,Belle:2010ldr,LHCb:2014scu,LHCb:2012ihl,LHCb:2023fqx,LHCb:2017zar,LHCb:2020cyw,LHCb:2020ist,LHCb:2024rto,LHCb:2024wnx,LHCb:2025rfy}.
The $B_{s}\to K^{-}\pi^{+}$ is an analog of $B_{s}\to D_{s}^{-}D^{+}$,
where the $c$ quark is replaced by the quark of a same type $c\to u$.
For other processes such analogs are not measured (Cabibbo suppressed),
but we report their numbers since they follow straightforwardly from
our calculations. Namely we give results for
\[
\begin{array}{ll}
B_{s}\to K^{-}D^{+} & \quad c\leftrightarrow u\text{ in }B_{s}\to D_{s}^{-}\pi^{+},\\
B_{s}\to K^{-}D^{*+} & \quad c\leftrightarrow u\text{ in }B_{s}\to D_{s}^{-}\rho^{+},\\
B_{s}\to\phi D^{0} & \quad c\leftrightarrow u\text{ in }B_{s}\to\phi\bar{D^{0}},\\
B_{s}\to\phi\pi & \quad c\to u\text{ in }B_{s}\to\phi\eta_{c}\text{ (factor }\nicefrac{1}{2}\text{)},
\end{array}
\]
where a factor $\nicefrac{1}{2}$ appears in the last case to reflect
the pion quark current structure, where $u$ is mixed with $d$. We
do not reevaluate all other processes addressed in publications mentioned
above since we are confident about our model providing consistent
results over the time. Nevertheless, some model parameters have slightly
shifted, thus we compute $B_{s}\to\phi\:J/\psi$ to compare and cross
check.

Besides the ``new data'' argument, other motivations are present.
The performed investigation is a broad test of our weak sector understanding
in the Standard Model (SM) and of the CCQM model: predictions we give
(as whole) depend on eight Cabibbo-Kobayashi-Maskawa (CKM) matrix
elements (all except $V_{td}$) and predict the relative importance
of different flavour-topology amplitudes. Particularly, $B_{s}$ measurements
are complementary to $B_{d}$ ones: they both have dynamics dominated
by a singe heavy quark and their comparison may give more insight
into penguin contributions and flavor changing neutral current searches.
All this has then implications for the parameter determination in
the SM and in other theoretical approaches (such as heavy quark effective
theory) or for a possible new physics (NP) discovery. The existence
of various models on the market with different treatments of hadronic
effects is useful to discriminate between the model dependence and
the actual physics: if an effect appears throughout most models it
may be model independent. In this sense we see our study as relevant.

The processes (\ref{eq:Processes}) have been theoretically addresses
in several contexts. The $B_{s}\to D_{s}^{-}\pi^{+}$ and $B_{s}\to D_{s}^{-}\rho^{+}$
reactions get neither penguin nor weak-annihilation contributions
and are therefore generally considered as theoretically clean non-leptonic
processes. The former serves as a normalization channel for various
other nonleptonic $B_{s}$ decay modes with a single $c$ quark in
the final state, from where the importance to understand it experimentally
and theoretically. It was addressed in \cite{Fleischer:2010ca} and
\cite{Piscopo:2023opf} to study the factorization, in \cite{Cai:2021mlt}
to probe NP and in other sources too. Explicit numbers for the $D_{s}^{-}\rho^{+}$
final state were given in \cite{Deandrea:1993ma} within the factorization
approximation.

The two-charm meson decay $B_{s}\to D_{s}^{-}D^{+}$ can be also used
to investigate the factorization: from the comparison to analogous
beauty baryon decays the importance of non-factorizable effects can
be deduced \cite{LHCb:2014scu}. In addition, when appropriate clean
observables are defined, this decay channel may serve for NP searches
too, see e.g. \cite{Jung:2014jfa}. As already mentioned earlier,
the $B_{s}\to K^{-}\pi^{+}$ process is an analog of the latter obtained
by $c\to u$. The theory of charmless $B_{s}$ decays to light states
was addressed multiple times, e.g. in \cite{Ali:2007ff,Cheng:2009mu},
where authors compute branching fractions, CP-violating asymmetries
and discuss NP effects, or in \cite{Gronau:2000md} where the subject
is the weak phase $\gamma\equiv\text{Arg}\left(V_{ub}^{*}\right)$
determination. The reference \cite{Yang:2025chl} then focuses on
further corrections (to mesonic wave function and distribution amplitudes)
of this decay.

The $B_{s}\to\phi\bar{D^{0}}$ process (see e.g. \cite{Zhou:2015jba,Talebtash:2016tym,Bakhshi:2024hds})
is generally seen as complementary to other $B_{s}$ and $B_{d}$
decays, possibly improving constraints on various weak physics parameters.

The last non-leptonic process we consider is $B_{s}\to\phi\eta_{c}$.
The authors of the reference \cite{Ke:2017wni} use this decay to
determine the $B_{s}\to\phi$ matrix element, then argue about the
implications for $f\left(980\right)$, which they claim to be a tetraquark.
The work \cite{Xiao:2019mpm} investigates various $B_{s}\to\eta_{c}h$
transitions including known next-to-leading order (NLO) contributions.
The authors demonstrate that these have an important effect. 

Semileptonic decays are significantly more theoretically clean and
consequently more precisely predicted and more discussed. The decays
$B_{s}\to D_{s}^{-}\mu^{+}\nu_{\mu}$ and $B_{s}\to K^{-}\mu^{+}\nu_{\mu}$
allow do determine $V_{cb}$ and $V_{ub}$ elements and their ratio
from exclusive processes and in this sense they constraint NP searches.
They provide complementary information to $B_{d}$ processes, the
presence of the heavier valence quark $s$ may benefit to lattice
quantum chromodynamics (QCD) calculations by providing smaller statistical
errors and finite-volume effects \cite{McLean:2019sds}. A large body
of literature dealing with these decays can be cited, such as \cite{Atoui:2013zza,McLean:2019qcx,Bordone:2019guc,Martinelli:2022xir,Cui:2023jiw}
and \cite{Bouchard:2014ypa,Flynn:2015mha,Gonzalez-Solis:2021pyh,Martinelli:2022xir,Flynn:2023nhi},
with the lattice QCD starting to dominate in theoretical evaluations.

The decay $B_{s}\to\phi^{0}e^{+}e^{-}$ is a $\bar{b}\to\bar{s}$
flavor-changing neutral current process which, in the SM, proceeds
only via loops. This makes of it a natural candidate for NP searches:
hypothetical new particles can enter the loop and modify the amplitude.
The muon version was measured already a time ago \cite{CDF:2011grz}
and some theory-experiment tensions appeared \cite{LHCb:2015wdu}
(true also for $B\to K^{(*)}\mu^{+}\mu^{-}$). The $e^{+}e^{-}$ decay
represents a consistency test with respect to its $\mu^{+}\mu^{-}$
counterpart and, on the same footing, is interesting as NP and lepton
universality probe. Most literature does not differentiate and works
with a general dilepton $\ell\bar{\ell}$ in the final state, e.g.
\cite{Bharucha:2015bzk}.

As stated already, we use the covariant confined quark model as a
tool to describe processes which include hadrons. The CCQM is an enhanced
effective-theory model that unifies covariance, gauge interactions
and quark confinement while keeping the number of free parameters
limited. It is based on a non-local interaction Lagrangian where quark-hadron
interaction vertex is introduced, allowing a hadron to turn into its
constituent quarks and vice versa. Then, on the quark level, we work
with effective four fermion operators. Acknowledging the impressive
progress in methods with small model dependence (in particular the
lattice QCD), we consider quark models as a more versatile and valid
complements. We describe the CCQM in more detail later in a dedicated
section.

The text is structured as follows: first, we present the general theory
of weak decays formulated in terms of effective Hamiltonians, hadronic
form factors and decay constants. Next, we present the CCQM and its
basic features, coming next to our results and conclusion. The theoretical
content we need to cover is large, therefore we keep out text concise
and provide only essential information with numerous references to
our previous works, where details can be found.

\section{Theory}

\subsection{Weak decays in effective theory}

Weak decays of hadrons proceed, on the quark level, through weak decays
of valence quarks. The transition from the initial state fermions
to the final state fermions involves various Feynman diagrams which
however need not to be reevaluated for every calculation. Effectively,
one can define a small number of (four-fermion) operators and weight
their action by (scale dependent) Wilson coefficients. A given operator
integrates contributions from various diagrams and several theoretical
groups dedicate their work to evaluate them and find the value of
the corresponding Wilson coefficient. The number of effective operators
is given by the Lorentz, chiral and color structure of the connected
particles (fermions). The choice of operators depends on the underlying
process and we will not list operator sets for all cases we address,
because this is a well-know model-independent framework described
in details in \cite{Buchalla:1995vs}. Nevertheless it seems appropriate
to illustrate the effective approach on a model example. We choose
the $B_{s}\to\phi^{0}e^{+}e^{-}$ decay which is rich in contributing
operators. For this process the effective Hamiltonian is written 
\begin{equation}
B_{s}\to\phi^{0}e^{+}e^{-}:\qquad\mathcal{H}_{\text{eff.}}=-\frac{4G_{F}}{\sqrt{2}}\lambda_{t}\sum_{i=1}^{10}C_{i}\left(\mu\right)\mathcal{O}_{i}\left(\mu\right)\label{eq:EffHam}
\end{equation}
with
\begin{alignat}{2}
\mathcal{O}_{1} & =\left(\bar{s}_{a_{1}}\gamma^{\mu}P_{L}c_{a_{2}}\right)\left(\bar{c}_{a_{2}}\gamma_{\mu}P_{L}b_{a_{1}}\right),\quad &  & \mathcal{O}_{2}=\left(\bar{s}\gamma^{\mu}P_{L}c\right)\left(\bar{c}\gamma_{\mu}P_{L}b\right),\nonumber \\
\mathcal{O}_{3} & =\left(\bar{s}\gamma^{\mu}P_{L}b\right)\sum_{q}\left(\bar{q}\gamma_{\mu}P_{L}q\right), &  & \mathcal{O}_{4}=\left(\bar{s}_{a_{1}}\gamma^{\mu}P_{L}b_{a_{2}}\right)\sum_{q}\left(\bar{q}_{a_{2}}\gamma_{\mu}P_{L}q_{a_{1}}\right),\nonumber \\
\mathcal{O}_{5} & =\left(\bar{s}\gamma^{\mu}P_{L}b\right)\sum_{q}\left(\bar{q}\gamma_{\mu}P_{R}q\right), &  & \mathcal{O}_{6}=\left(\bar{s}_{a_{1}}\gamma^{\mu}P_{L}c_{a_{2}}\right)\sum_{q}\left(\bar{q}_{a_{2}}\gamma_{\mu}P_{R}q_{a_{1}}\right),\label{eq:model_WC}\\
\mathcal{O}_{7} & =\frac{e}{16\pi^{2}}\bar{m_{b}}\left(\bar{s}\sigma^{\mu\nu}P_{R}b\right)F_{\mu\nu}, &  & \mathcal{O}_{8}=\frac{g}{16\pi^{2}}\bar{m_{b}}\left(\bar{s}\sigma^{\mu\nu}P_{R}\mathbf{T}_{a_{1}a_{2}}b_{a_{2}}\right)\mathbf{G}_{\mu\nu},\nonumber \\
\mathcal{O}_{9} & =\frac{e}{16\pi^{2}}\left(\bar{s}\gamma^{\mu}P_{L}b\right)\left(\bar{\ell}\gamma_{\mu}\ell\right), &  & \mathcal{O}_{10}=\frac{e}{16\pi^{2}}\left(\bar{s}\gamma^{\mu}P_{L}b\right)\left(\bar{\ell}\gamma_{\mu}\gamma_{5}\ell\right).\nonumber 
\end{alignat}
The symbol $G_{F}$ denotes the Fermi coupling constant, $\lambda_{t}$
the product of the CKM elements $\lambda_{t}=|V_{tb}V_{ts}^{\ast}|$,
$\sigma^{\mu\nu}=\frac{i}{2}\left[\gamma^{\mu},\gamma^{\nu}\right]$,
$C_{i}\left(\mu\right)$ are renormalization-scale $\mu$ dependent
Wilson coefficients and $\mathcal{O}_{i}$ the effective operators.
These are built from quark fields ($s$, $c$, $b$ and $q$) which
carry color subscript $a_{i}$, shown only for nontrivial combinations.
Further $P_{L/R}=\left(1\mp\gamma_{5}\right)/2$, $\bar{m_{b}}$ is
the QCD bottom quark mass (different from the constituent quark mass
of the CCQM), $F_{\mu\nu}$ is the electromagnetic tensor, $\mathbf{G}_{\mu\nu}$
the gluon field strength tensor and $\mathbf{T}_{a_{1}a_{2}}$ are
the generators of the color $SU\left(3\right)$ group. Operators $\mathcal{O}_{1,2}$
are referred to as current-current, $\mathcal{O}_{3-6}$ as QCD penguins,
$\mathcal{O}_{7,8}$ are labeled ``magnetic penguin'' and $\mathcal{O}_{9,10}$
as electroweak penguins. All these are then used to express the decay
amplitude through Wilson coefficients and transition amplitudes $B_{s}\to\phi$.
Using hadonic form factors to parameterize the latter, one gets the
final formula for the amplitude and the decay with. We avoid writing
the expression here because of its length, the reader can find it
in references \cite{Misiak:1992bc,Asatryan:2001zw,Dubnicka:2016nyy}.

\subsection{Hadronic form factors}

Assuming the na\"{i}ve factorization, the amplitude splits into the
product of a hadronic transition amplitude and a leptonic decay constant
(for hadronic decays) or a leptonic vector (for semi-leptonic decays).
The transition amplitude is then parameterized by form factors which
appear as scalar $q^{2}$-dependent coefficients in front of independent
Lorentz structures. With $B_{s}$ a pseudoscalar particle (denoted
as $P$, a vector particle as $V$), we deal only with a small number
of hadronic transition amplitudes. The definition of form factors
then stands ($P=p_{1}+p_{2}$, $q=p_{1}-p_{2}$)
\begin{align}
 & \left(P'P\right)_{V}=\left\langle P'_{\left[\bar{q_{3}}q_{2}\right]}(p_{2})|\bar{q}_{2}\gamma^{\mu}q_{1}|P_{\left[\bar{q_{3}}q_{1}\right]}(p_{1})\right\rangle =F_{+}(q^{2})P^{\mu}+F_{-}(q^{2})q^{\mu},\nonumber \\
 & \left(P'P\right)_{S}=\left\langle P'_{\left[\bar{q_{3}}q_{2}\right]}(p_{2})|\bar{q}_{2}q_{1}|P_{\left[\bar{q_{3}}q_{1}\right]}(p_{1})\right\rangle =\left(m_{P'}+m_{P}\right)F_{S}(q^{2}),\nonumber \\
 & \left(VP\right)_{V}=\left\langle V_{\left[\bar{q_{3}}q_{2}\right]}(p_{2},\epsilon)|\bar{q}_{2}\gamma^{\mu}\left(1-\gamma_{5}\right)q_{1}|P_{\left[\bar{q_{3}}q_{1}\right]}(p_{1})\right\rangle =\frac{\epsilon_{\nu}^{\dagger}}{m_{P}+m_{V}}\label{eq:FFsDef}\\
 & \quad\times\left[-g^{\mu\nu}(P\cdot q)\,A_{0}(q^{2})+P^{\mu}P^{\nu}A_{+}(q^{2})+q^{\mu}P^{\nu}A_{-}(q^{2})+i\varepsilon^{\mu\nu\alpha\beta}P_{\alpha}q_{\beta}V(q^{2})\right],\nonumber \\
 & \left(VP\right)_{T}=\left\langle V_{\left[\bar{q_{3}}q_{2}\right]}(p_{2},\epsilon)|\bar{q}_{2}\sigma^{\mu\nu}q_{\nu}(1+\gamma_{5})q_{1}|P_{\left[\bar{q_{3}}q_{1}\right]}(p_{1})\right\rangle =\epsilon_{\nu}^{\dagger}\left[-\left(g^{\mu\nu}-\frac{q^{\mu}q^{\nu}}{q^{2}}\right)\right.\nonumber \\
 & \quad\left.\times(P\cdot q)\,a_{0}(q^{2})+\left(P^{\mu}P^{\nu}-\frac{q^{\mu}P^{\nu}(P\cdot q)}{q^{2}}\right)\,a_{+}(q^{2})+i\varepsilon^{\mu\nu\alpha\beta}P_{\alpha}q_{\beta}\,q(q^{2})\right],\nonumber 
\end{align}
where we list form factors relevant for our calculations. To complete
the model example from the previous section, the $B_{s}\to\phi^{0}e^{+}e^{-}$
matrix element is expressed as
\begin{align*}
\mathcal{M} & =\frac{G_{F}}{\sqrt{2}}\frac{\alpha\lambda_{t}}{2\pi}\left\{ C_{9}^{\text{eff.}}\mathcal{M}_{1}^{\mu}\left(\bar{e}\gamma_{\mu}e\right)-\frac{2\bar{m}_{b}}{q^{2}}C_{7}^{\text{eff.}}\mathcal{M}_{2}^{\mu}\left(\bar{e}\gamma_{\mu}e\right)+C_{10}\mathcal{M}_{1}^{\mu}\left(\bar{e}\gamma_{\mu}\gamma_{5}e\right)\right\} ,\\
\mathcal{M}_{1}^{\mu} & =\left\langle \phi|\bar{s}\gamma^{\mu}\left(1-\gamma_{5}\right)b|B_{s}\right\rangle ,\\
\mathcal{M}_{2}^{\mu} & =\left\langle \phi|\bar{s}i\sigma^{\mu\nu}q_{\nu}(1+\gamma_{5})b|B_{s}\right\rangle ,
\end{align*}
with
\begin{align*}
C_{7}^{\text{eff.}} & =C_{7}^{\text{eff.}}\left(C_{5},C_{6},C_{7}\right),\\
C_{9}^{\text{eff.}} & =C_{9}^{\text{eff.}}\left(C_{1},C_{2},C_{3},C_{4},C_{5},C_{6},C_{9}\right)
\end{align*}
being some known, specific function of Wilson coefficients from (\ref{eq:EffHam}),
see \cite{Dubnicka:2016nyy}, and $\alpha$ is the electromagnetic
(EM) coupling. The amplitudes $\mathcal{M}_{1,2}^{\mu}$ are parameterized
as given by (\ref{eq:FFsDef}). 

To express the decay widths we use intermediate objects, the helicity
form factors, which are defined as 
\begin{align*}
P\to P:\\
H_{t}= & \,\frac{1}{\sqrt{2}}\left[\left(m_{P}^{2}-m_{P'}^{2}\right)F_{+}+q^{2}F_{-}\right],\\
H_{0}= & \,\frac{2m_{P}\left|\boldsymbol{p_{2}}\right|}{\sqrt{q^{2}}}F_{+},\\
P\to V:\\
H_{t}= & \,\frac{1}{m_{P}+m_{P'}}\frac{m_{P}}{m_{P'}}\frac{\left|\boldsymbol{p_{2}}\right|}{\sqrt{q^{2}}}\left[\left(m_{P}^{2}-m_{P'}^{2}\right)\left(A_{+}-A_{0}\right)+q^{2}A_{-}\right],\\
H_{\pm}= & \,\frac{1}{m_{P}+m_{P'}}\left[-\left(m_{P}^{2}-m_{P'}^{2}\right)A_{0}\pm2m_{P}\left|\boldsymbol{p_{2}}\right|V\right],\\
H_{0}= & \,\frac{1}{m_{P}+m_{P'}}\frac{1}{2m_{P'}\sqrt{q^{2}}}\left[-\left(m_{P}^{2}-m_{P'}^{2}\right)\left(m_{P}^{2}-m_{P'}^{2}-q^{2}\right)A_{0}\right.\\
 & \left.+4m_{P}^{2}\left|\boldsymbol{p_{2}}\right|^{2}A_{+}\right],
\end{align*}
where $\left|\boldsymbol{p_{2}}\right|$ is the absolute value of
the three-momentum of outgoing particles in the $B_{s}$ rest frame
and can be expressed through the K\"{a}ll\'{e}n $\lambda$ function
$\left|\boldsymbol{p_{2}}\right|=\sqrt{\lambda(m_{P}^{2},m_{P'}^{2},q^{2})}/(2m_{P})$.
The $P\to V$ tensor form factors $a_{0}$, $a_{+}$ and $q$ appear
only for $B_{s}\to\phi^{0}e^{+}e^{-}$, we comment on them in the
result section.

\subsection{Decay widths}

We provide here the list of decay-width formulas, all of which can
be found in the literature or our previous publications (see e.g.
Appendix A in \cite{Ivanov:2006ni}). For each formula we list the
relevant processes and show Feynman diagrams. When dealing with hadronic
decays we refer to different diagram topologies as class 1 (color
allowed) or class 2 (color suppressed). Letters $a_{i}$ and $C_{X}^{(\text{eff.})}$
denote combinations of the Wilson coefficients
\begin{align*}
a_{1} & =C_{2}+\xi C_{1},\\
a_{2} & =C_{1}+\xi C_{2},\\
C_{2}^{\text{eff.}} & =C_{2}+\xi C_{1}+C_{4}+\xi C_{3},\\
C_{6}^{\text{eff.}} & =C_{6}+\xi C_{5},\\
C_{P} & =C_{1}+\xi C_{2}+C_{3}+\xi C_{4}-C_{5}-\xi C_{6},\\
C_{V} & =C_{1}+\xi C_{2}+C_{3}+\xi C_{4}+C_{5}+\xi C_{6},
\end{align*}
where $\xi=1/N_{c}$ with $N_{c}$ being the number of colors and
$f_{M}$ denotes the leptonic decay constant of the meson $M$ ($f_{M}^{S}$
is the scalar decay constant related to the scalar form factor $F_{S}$).

\subsubsection{Class 1, $PS\to PS+PS$}

\noindent Included decays: $B_{s}^{0}\to D_{s}^{-}\pi^{+}$

\noindent Contributing operators: $\mathcal{O}_{1},\mathcal{O}_{2}$

\noindent Decay width formula:
\[
\Gamma(B_{s}^{0}\to D_{s}^{-}\pi^{+})=\frac{1}{16\pi m_{B_{s}}^{2}}\left|\mathbf{p}_{2}\right|G_{F}^{2}\left(V_{cb}V_{ud}\right)^{2}\left[a_{1}f_{\pi}m_{\pi}H_{t}^{B_{s}\to D_{s}}(m_{\pi}^{2})\right]^{2}
\]

\begin{figure}
\begin{centering}
\includegraphics[width=0.45\textwidth]{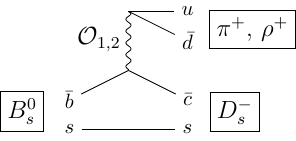}
\par\end{centering}
\caption{The feynman diagram of decays $B_{s}^{0}\to D_{s}^{-}+\pi^{+}$ and
$B_{s}^{0}\to D_{s}^{-}+\rho^{+}$.}
\label{Fig_Dec_1}

\end{figure}

\noindent Diagram: Figure \ref{Fig_Dec_1}. 

\subsubsection{Class 1, $PS\to PS+V$}

\noindent Included decays: $B_{s}^{0}\to D_{s}^{-}\rho^{+}$

\noindent Contributing operators: $\mathcal{O}_{1},\mathcal{O}_{2}$

\noindent Decay width formula:
\[
\Gamma(B_{s}^{0}\to D_{s}^{-}\rho^{+})=\frac{1}{16\pi m_{B_{s}}^{2}}\left|\mathbf{p}_{2}\right|G_{F}^{2}\left(V_{cb}V_{ud}\right)^{2}\left[a_{1}f_{\rho}m_{\rho}H_{0}^{B_{s}\to D_{s}}(m_{\rho}^{2})\right]^{2}
\]

\noindent Diagram: Figure \ref{Fig_Dec_1}.

\subsubsection{Class 1 + penguin, $PS\to PS+PS$}

\noindent Included decays: $B_{s}^{0}\to D_{s}^{-}D^{+}$, $B_{s}^{0}\to K^{-}\pi^{+}$

\noindent Contributing operators: $\mathcal{O}_{1},\dots,\mathcal{O}_{6}$

\noindent Decay width formula:
\begin{align*}
\Gamma(B_{s}^{0}\to D_{s}^{-}D^{+})= & \frac{1}{16\pi m_{B_{s}}^{2}}\left|\mathbf{p}_{2}\right|G_{F}^{2}\left(V_{cb}V_{cd}\right)^{2}\\
 & \times\left[C_{2}^{\text{eff}}f_{D}m_{D}H_{t}^{B_{s}\to D_{s}}(m_{D}^{2})+2C_{6}^{\text{eff}}f_{D}^{S}F_{S}^{B_{s}\to D_{s}}(m_{D}^{2})\right]^{2}
\end{align*}

\begin{figure}
\begin{centering}
\includegraphics[width=0.45\textwidth]{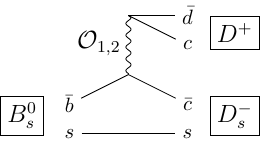}\quad{}\includegraphics[width=0.45\textwidth]{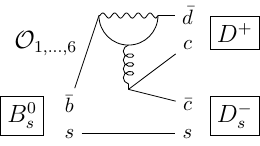}
\par\end{centering}
\caption{The Feynman diagrams of the decay $B_{s}^{0}\to D_{s}^{-}+D^{+}$.}
\label{Fig_Dec_2}
\end{figure}

\noindent Diagram: Figure \ref{Fig_Dec_2}.

\subsubsection{Class 2, $PS\to V+PS$}

\noindent Included decays: $B_{s}^{0}\to\phi\overline{D^{0}}$

\noindent Contributing operators: $\mathcal{O}_{1},\mathcal{O}_{2}$

\noindent Decay width formula:
\[
\Gamma(B_{s}^{0}\to\phi\overline{D^{0}})=\frac{1}{16\pi m_{B_{s}}^{2}}\left|\mathbf{p}_{2}\right|G_{F}^{2}\left(V_{cb}V_{us}\right)^{2}\left[a_{2}f_{\overline{D^{0}}}m_{\overline{D^{0}}}H_{t}^{B_{s}\to\phi}(m_{\overline{D^{0}}}^{2})\right]^{2}
\]

\begin{figure}
\begin{centering}
\includegraphics[width=0.45\textwidth]{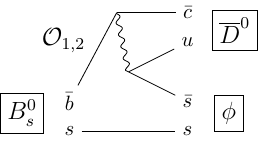}
\par\end{centering}
\caption{The Feynman diagram of the decay $B_{s}^{0}\to\phi+\overline{D^{0}}$.}
\label{Fig_Dec_3}
\end{figure}

\noindent Diagram: Figure \ref{Fig_Dec_3}.

\subsubsection{Class 2 + penguin, $PS\to V+PS$}

\noindent Included decays: $B_{s}^{0}\to\phi\eta_{c}$, $B_{s}^{0}\to\phi\pi$

\noindent (additional factor $1/2$ in the decay width with the pion)

\noindent Contributing operators: $\mathcal{O}_{1},\dots,\mathcal{O}_{6}$

\noindent Decay width formula:
\[
\Gamma(B_{s}^{0}\to\phi\eta_{c})=\frac{G_{F}^{2}}{16\pi}\frac{\left|\mathbf{p}_{2}\right|}{m_{B_{s}}^{2}}\left|V_{cb}V_{cs}^{\dagger}\right|^{2}\left[C_{P}f_{\eta_{c}}m_{\eta_{c}}H_{t}^{B_{s}\to\phi}\left(m_{\eta_{c}}^{2}\right)\right]^{2}
\]

\begin{figure}
\begin{centering}
\includegraphics[width=0.45\textwidth]{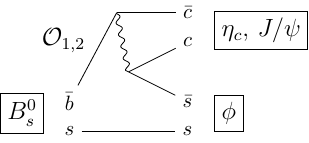}\quad{}\includegraphics[width=0.45\textwidth]{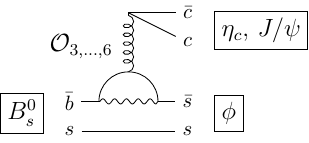}
\par\end{centering}
\caption{The Feynman diagrams of decays $B_{s}^{0}\to\phi+\eta_{c}$ and $B_{s}^{0}\to\phi+J/\psi$.}
\label{Fig_Dec_4}
\end{figure}

\noindent Diagram: Figure \ref{Fig_Dec_4}.

\subsubsection{Class 2 + penguin, $PS\to V+V$}

\noindent Included decays: $B_{s}^{0}\to\phi+J/\psi$

\noindent Contributing operators: $\mathcal{O}_{1},\dots,\mathcal{O}_{6}$

\noindent Decay width formula:
\[
\Gamma(B_{s}^{0}\to\phi J/\psi)=\frac{G_{F}^{2}}{16\pi}\frac{\left|\mathbf{p}_{2}\right|}{m_{B_{s}}^{2}}\left|V_{cb}V_{cs}^{\dagger}\right|^{2}C_{V}^{2}f_{J/\psi}^{2}m_{J/\psi}^{2}\sum_{i=0,\pm}\left[H_{i}^{B_{s}\to\phi}\left(m_{J/\psi}^{2}\right)\right]^{2}
\]

\noindent Diagram: Figure \ref{Fig_Dec_4}.

\subsubsection{Semileptonic, with neutrino, $PS\to PS+\mu^{+}\nu$}

\noindent Included decays: $B_{s}^{0}\to D_{s}^{-}\mu^{+}\nu$, $B_{s}^{0}\to K^{-}\mu^{+}\nu$

\noindent Decay width formula:
\begin{align*}
\Gamma(B_{s}^{0} & \to D_{s}^{-}\mu^{+}\nu)=\frac{G_{F}^{2}|V_{cb}|^{2}}{\left(2\pi\right)^{3}}\\
 & \times\int_{m_{\mu}}^{(m_{B_{s}}-m_{D_{s}})^{2}}\frac{(q^{2}-m_{\mu}^{2})}{12m_{B_{s}}^{2}q^{2}}\left|\mathbf{p}_{2}\right|\left[\left(1+\frac{m_{\mu}^{2}}{2q^{2}}\right)H_{0}^{2}+\frac{3m_{\mu}^{2}}{2q^{2}}H_{t}^{2}\right]dq^{2}
\end{align*}

\begin{figure}
\begin{centering}
\includegraphics[width=0.45\textwidth]{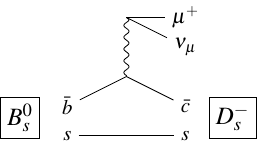}\quad{}\includegraphics[width=0.45\textwidth]{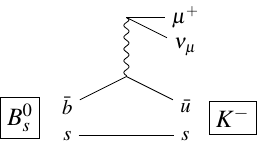}
\par\end{centering}
\caption{The Feynman diagrams of decays $B_{s}^{0}\to D_{s}^{-}+\mu^{+}\nu$
and $B_{s}^{0}\to K^{-}+\mu^{+}\nu$.}
\label{Fig_Dec_6}
\end{figure}

\noindent Diagram: Figure \ref{Fig_Dec_6}.

\subsubsection{Semileptonic, with dilepton, $PS\to V+\ell^{+}\ell^{-}$}

\noindent Included decays: $B_{s}^{0}\to\phi^{0}e^{+}e^{-}$

\noindent Decay width formula:
\begin{align*}
\Gamma(B_{s}^{0}\to\phi^{0}e^{+}e^{-}) & =\int dq^{2}\frac{G_{F}^{2}}{\left(2\pi\right)^{3}}\frac{\alpha^{2}|V_{tb}V_{ts}^{*}|^{2}}{(2\pi)^{2}}\frac{\left|\mathbf{p}_{2}\right|q^{2}\beta_{e}}{12m_{B_{s}}^{2}}H_{\text{tot}},
\end{align*}
where $\beta_{e}=\sqrt{1-4m_{e}^{2}/q^{2}}$ and $H_{\text{tot}}$
is function of invariant form factors and Wilson coefficients $H_{\text{tot}}=H_{\text{tot}}(A_{\pm},A_{0},V,a_{0},a_{+},g,C_{1,\dots,10})$
whose explicit form can be found in \cite{Dubnicka:2016nyy}. 

\begin{figure}
\begin{centering}
\includegraphics[width=0.45\textwidth]{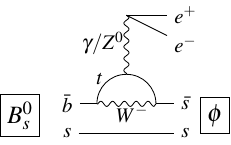}
\par\end{centering}
\caption{The Feynman diagram of the decay $B_{s}^{0}\to\phi^{0}e^{+}e^{-}$.}
\label{Fig_Dec_7}
\end{figure}

\noindent Diagram: Figure \ref{Fig_Dec_7}.

\section{Model}

\subsection{Formulation}\label{subsec:Formulation}

The role of the CCQM model is to predict the invariant form factors
and leptonic decay widths. These objects enclose also contributions
from non-perturbative hadronic effects related to $B_{s}$ not being
elementary, but being rather a bound state of strongly interacting
quarks. The model is an effective theory approach that uses a non-local
quark-hadron interaction Lagrangian and has a straightforward interpretation:
the introduced interaction allows a hadron to turn into its constituent
quarks which then interact according to the SM rules. The final state
quarks are combined again to form outgoing hadrons. The model is constructed
for various multiquark states, we use the meson formulation
\begin{align*}
{\cal L}_{{\rm int}}^{\mathrm{CCQM}}(x) & =g_{M}\cdot M_{\left(\mu\right)}(x)\cdot J_{M}^{\left(\mu\right)}(x),\\
J_{M}^{\left(\mu\right)}(x) & =\int\!\!dx_{1}\!\!\int\!\!dx_{2}\,F_{M}(x,x_{1},x_{2})\cdot\bar{q}_{f_{1}}^{a}(x_{1})\,\Gamma_{M}^{\left(\mu\right)}\,q_{f_{2}}^{a}(x_{2}).
\end{align*}
Here the mesonic field $M$ interacts with the quark current $J_{M}$
with strength $g_{M}$. The symbol $\Gamma_{M}$ represents an appropriate
string of Dirac matrices to describe the spin of $M$ and may carry
a Lorentz index. Subscripts $f$ denote flavors and superscripts $a$
the color. We construct the vertex function $F$ as
\begin{align*}
F_{M}(x,x_{1},x_{2}) & =\delta(x-w_{1}x_{1}-w_{2}x_{2})\Phi_{M}\Big[(x_{1}-x_{2})^{2}\Big],\\
w_{i} & =m_{i}/(m_{1}+m_{2}),
\end{align*}
so that $F$ is fully covariant and the hadron is positioned at the
quark system's barycenter. The definition of $\Phi_{M}$ stands
\begin{equation}
\widetilde{\Phi}_{M}(-\,k^{2})=\exp\left(k^{2}/\Lambda_{M}^{2}\right)\label{eq:verFnForm}
\end{equation}
and is guided also by the computational convenience. The symbol $\widetilde{\Phi}$
denotes the Fourier transform into the momentum variable and the minus
sign indicates that the Wick-rotated expression in the Euclidean space
falls off quickly for integrals to be finite. The letter $\Lambda$
represents a free model parameter linked to the meson $M$. Additional
model parameters are constituent quark masses ($m_{q}=m_{u}=m_{d}$,
$m_{s}$, $m_{c}$, $m_{b}$) and a cutoff parameter $\lambda^{\text{cutoff}}$
needed to implement the confinement (see later).

\begin{figure}
\begin{centering}
\includegraphics[width=1\textwidth]{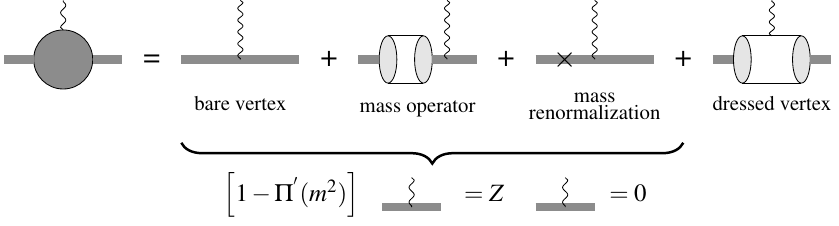}
\par\end{centering}
\caption{Diagrammatic representation of the compositeness condition.}\label{fig:diagramCompositeness}

\end{figure}

There are few important principles that the CCQM incorporates in addition
to the Lagrangian. First, as is evident from the construction above,
mesonic and quark fields are present, both elementary. If the two
fields can appear inside Feynman diagrams one worries about a possible
double counting. Indeed, in nature, mesons and quarks are not independent
but former are built from latter. We address this concern in a way
described in \cite{Salam:1962ap,Weinberg:1962hj,Issadykov:2015iba},
meaning that we apply the so-called compositeness condition
\begin{equation}
Z_{M}^{1/2}=1-3g_{M}^{2}\Pi_{M}^{'}(m_{M}^{2})/(4\pi)=0,\label{eq:compositeness}
\end{equation}
where $Z_{M}^{1/2}$ is the meson renormalization constant and $\Pi_{M}^{'}$
is, for scalar mesons, the derivative of the meson mass operator.
For vector mesons it is the derivative of the corresponding scalar
par. Diagrammatically is (\ref{eq:compositeness}) expressed in Fig.
\ref{fig:diagramCompositeness}. 
 The condition implies that the physical state has no overlap with
the bare state, the interaction involves only constituents (quarks),
which are however always virtual and therefore excluded from the space
of physical parameters. The condition is satisfied by tuning the value
of $g_{M}$. More details can be found for example in the section
II of \cite{Ivanov:2011aa} and references therein. 

An additional feature of the CCQM is the quark confinement that extends
the model applicability to a hadron, whose mass exceeds the masses
of its constituents summed. The approach is inspired by the confined-propagator
technique which writes down the quark propagator in the Schwinger
representation and introduces a cut in the integration limit
\[
\frac{1}{m^{2}-p^{2}}=\int_{0}^{\infty}d\alpha\,e^{-\alpha(m^{2}-p^{2})}\longrightarrow\int_{0}^{1/\lambda^{2}}d\alpha\,e^{-\alpha(m^{2}-p^{2})}=\frac{1-e^{-\frac{m^{2}-p^{2}}{\lambda^{2}}}}{m^{2}-p^{2}}.
\]
The singularity for $p^{2}\to m^{2}$ is removed for the expression
on the right side, the propagator becomes an entire function without
poles. The lack of singularities indicates an absence of asymptotic
singe-particle quark states, thus quarks are effectively confined.
The procedure is for the CCQM more evolved (see Sec. II.c of \cite{Branz:2009cd}),
the technique is applied to the whole structure of the Feynman diagram
$F$. Having $n$ Schwinger parameters, one substitutes the unity
in the form $1=\int_{0}^{\infty}dt\,\delta(t-\sum_{i=1}^{n}\alpha_{i})$
to get a product of an integration over a simplex and one improper
integral 
\begin{align*}
\Pi & =\int_{0}^{\infty}d^{n}\alpha\,F\left(\alpha_{1},\dots,\alpha_{n}\right)\\
 & =\int_{0}^{\infty\to\frac{1}{\lambda^{2}}}dt\,t^{n-1}\int_{0}^{1}d^{n}\alpha\,\delta\left(t-\sum_{i=1}^{n}\alpha_{i}\right)F\left(t\alpha_{1},\dots,t\alpha_{n}\right).
\end{align*}
The cut is applied to the one-dimensional integral only, but still,
by analogous arguments, removes all existing thresholds in the quark-loop
diagrams and provides a quark confinement. The cutoff parameter $\lambda$
is a global model parameter.

Although we do not investigate radiative decays here, for completeness
we mention that the model includes also interactions with photons:
the free parts of the Lagrangian are gauged using the minimal subtraction
procedure, and the gauge-field exponential technique is applied to
the strong-interaction part, analogically to \cite{Terning:1991yt}.

\subsection{Application}

The computations proceed by evaluation of relevant Feynman diagrams.
The first step is the determination of the coupling $g_{M}$ for all
appearing mesons. This is done by applying the compositeness condition
(\ref{eq:compositeness}). Then one writes down the decay amplitude
which is expressed as a product of a decay constant (or leptonic vector)
and hadronic transition amplitue. The expression of the latter in
the CCQM reads 
\begin{align}
\left(P'P\right)_{V}= & N_{c}g_{P}g_{P'}\int\frac{d^{4}k}{(2\pi)^{4}i}\tilde{\Phi}_{P}[-(k+w_{13}p_{1})^{2}]\tilde{\Phi}_{P'}[-(k+w_{23}p_{2})^{2}]\nonumber \\
 & \quad\times\text{tr}\left[O^{\mu}S_{1}(k+p_{1})\gamma^{5}S_{3}(k)\gamma^{5}S_{2}(k+p_{2})\right],\nonumber \\
\left(P'P\right)_{S}= & N_{c}g_{P}g_{P'}\int\frac{d^{4}k}{(2\pi)^{4}i}\tilde{\Phi}_{P}[-(k+w_{13}p_{1})^{2}]\tilde{\Phi}_{P'}[-(k+w_{23}p_{2})^{2}]\nonumber \\
 & \quad\times\text{tr}\left[S_{1}(k+p)\gamma^{5}S_{3}(k)\gamma^{5}S_{2}(k+p_{2})\right],\label{eq:FFs_in_CCQM}\\
\left(VP\right)_{V}= & N_{c}g_{P}g_{V}\int\frac{d^{4}k}{(2\pi)^{4}i}\tilde{\Phi}_{P}[-(k+w_{13}p_{1})^{2}]\tilde{\Phi}_{V}[-(k+w_{23}p_{2})^{2}]\nonumber \\
 & \quad\times\text{tr}\left[O^{\mu}S_{1}(k+p_{1})\gamma^{5}S_{3}(k)\gamma^{\nu}(\epsilon_{2}^{\dagger})_{\nu}S_{2}(k+p_{2})\right],\nonumber \\
\left(VP\right)_{T}= & \quad N_{c}g_{P}g_{V}\int\frac{d^{4}k}{(2\pi)^{4}i}\tilde{\Phi}_{P}[-(k+w_{13}p_{1})^{2}]\tilde{\Phi}_{V}[-(k+w_{23}p_{2})^{2}]\nonumber \\
 & \quad\times\text{tr}\left[\sigma^{\mu\nu}q_{\nu}(1+\gamma^{5})S_{1}(k+p_{1})\gamma^{5}S_{3}(k)\gamma^{\nu}(\epsilon_{2}^{\dagger})_{\nu}S_{2}(k+p_{2})\right],\nonumber 
\end{align}
where $N_{c}$ is the number of colors, $O^{\mu}=\gamma^{\mu}(1-\gamma^{5})$,
$S$ denotes the quark propagator and $w_{ij}=m_{j}/(m_{i}+m_{j})$
with three contributing quarks. Indices $i=1,2$ label quarks in the
decaying meson and $i=2,3$ in the final state meson, with $i=3$
being the spectator quark. Once these expressions are evaluated and
compared to (\ref{eq:FFsDef}), the invariant form factors are extracted.
For what concerns the calculations of leptonic decay constants and
couplings $g_{M}$ (the previously mentioned compositeness condition),
they are similar and the corresponding CCQM expressions can be found
for example in \cite{Ivanov:2011aa} (formulas 5 and 15 there). The
above expressions are further processed: switching to the momentum
picture an exponential appears, it contains exponents from the vertex
function (\ref{eq:verFnForm}), from the Fourier-transformed propagators
and also from the Schwinger parametrization. This exponent is than
arranged by powers of $k$, formally written as $\exp(ak^{2}+2rk+z)$,
and the loop momenta integral is evaluated thanks to the identity
\[
\int d^{4}k\text{P}\left(k\right)e^{\left(ak^{2}+2rk+z\right)}=e^{z}\text{P}\left(\frac{1}{2}\frac{\partial}{\partial r}\right)\int d^{4}ke^{\left(ak^{2}+2rk\right)},
\]
where a know Gaussian integral appears. Parameters $a$, $r$ and
$z$ depend on $\Lambda_{P}$, $\Lambda_{P'}$( or $\Lambda_{V}$),
on $\alpha_{i}$ and on masses and external momenta, $\text{P}$ denotes
a polynomial from the trace evaluation. As a result, we get a polynomial
built from a differential operator acting on an exponential in variable
$r$. Again the computations are simplified using
\begin{equation}
\text{P}\left(\frac{1}{2}\frac{\partial}{\partial r}\right)e^{-\frac{r^{2}}{a}}=e^{-\frac{r^{2}}{a}}\text{P}\left(-\frac{r}{a}+\frac{1}{2}\frac{\partial}{\partial r}\right),\label{eq:commutation}
\end{equation}
where the commutation makes the operator act on a unity. We extensively
use the FORM \cite{Kuipers:2012rf} software, not only to evaluate
the trace, but also for formal manipulations with the differential
operator and commutations (\ref{eq:commutation}). The repeated use
of the chain rule removes all derivatives from the expression and
the integration over the Schwinger variables remains the last step
before getting results
\[
\Pi=\int_{0}^{\infty}d^{n}\alpha\,F\left(\alpha_{1},\dots,\alpha_{n}\right).
\]
Proceeding in accordance with the confinement implementation (Sec.
\ref{subsec:Formulation}), we perform a numerical integration using
a Java numerical library \cite{TorstenNahm}.

\section{Results}\label{sec:List-of-decays}

\subsection{Numerical inputs}

The values of Wilson coefficients evaluated at $\mu_{b}=4.8$ GeV
are taken from \cite{Descotes-Genon:2013vna}. They stand
\begin{align*}
C_{1} & =-0.2632,\quad C_{2}=1.0111,\quad C_{3}=-0.0055,\\
C_{4} & =-0.0806,\quad C_{5}=0.0004,\quad C_{6}=0.0009.
\end{align*}
We work in the large $N_{c}$ limit $\xi\to0$. The CCQM model inputs
are represented by constituent quark masses (in GeV)
\[
m_{u,d}=0.241,\quad m_{s}=0.428,\quad m_{c}=1.672,\quad m_{b}=5.046
\]
and they were established in previous works by fitting the data. The
same procedure was done for hadronic $\Lambda$ parameters which are
(in GeV)
\[
\Lambda_{B_{s}}=2.05,\;\Lambda_{D_{s}}=1.75,\;\Lambda_{\pi}=0.87,\;\Lambda_{\rho}=0.61,\;\Lambda_{\phi}=0.88,\;\Lambda_{\eta_{c}}=3.97.
\]
The cutoff parameter has a universal value $\lambda=0.181$ GeV. The
QCD quark masses are (in GeV) $\bar{m}_{c}=1.27$, $\bar{m}_{b}=4.68$
and $\bar{m}_{t}=173.3$ (they enter the $C_{9}^{\text{eff.}}$ coefficient
in semileptonic decays).

For reproducibility reasons let us state also values of nature constants.
We use (masses in GeV)
\begin{align*}
m_{B_{s}} & =5.366,\quad m_{D_{s}}=1.968,\quad m_{\phi}=1.019,\\
|V_{cb}| & =0.041,\quad|V_{ud}|=0.9737,\quad|V_{us}|=0.225,\\
|V_{cs}| & =0.987,\quad|V_{cd}|=0.224,\quad|V_{tb}V_{ts}^{*}|=0.041.
\end{align*}

\subsection{Derived quantities}

Using methods described earlier we derive, as basic outputs of the
CCQM, the leptonic decay constants and hadronic transition form factors.
The former are predicted to be (in GeV)
\begin{center}
\begin{tabular}{ll}
$f_{\pi}=0.130,$\vspace{0.2cm}
 & $\quad f_{\pi}^{S}=0.250,$\tabularnewline
$f_{\rho}=0.218,$\vspace{0.2cm}
 & $\quad f_{\eta_{c}}=0.627,$\tabularnewline
$f_{D}=0.206,$\vspace{0.2cm}
 & $\quad f_{D}^{S}=0.794,$\tabularnewline
\multicolumn{2}{l}{$\qquad f_{J/\psi}=0.415.$\vspace{0.2cm}
}\tabularnewline
\end{tabular}
\par\end{center}

\begin{figure}
\begin{centering}
\includegraphics[width=0.45\textwidth]{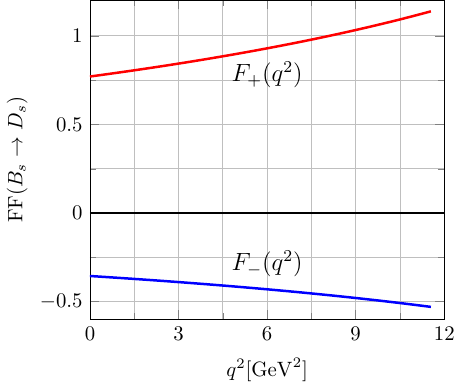}\quad{}\includegraphics[width=0.45\textwidth]{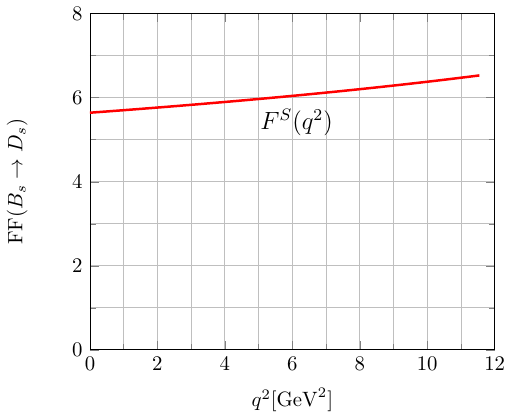}
\par\end{centering}
\caption{Form factors of the $B_{s}\to D_{s}$ transition.}
\label{Fig:FFs_D}

\end{figure}

\begin{figure}
\begin{centering}
\includegraphics[width=0.45\textwidth]{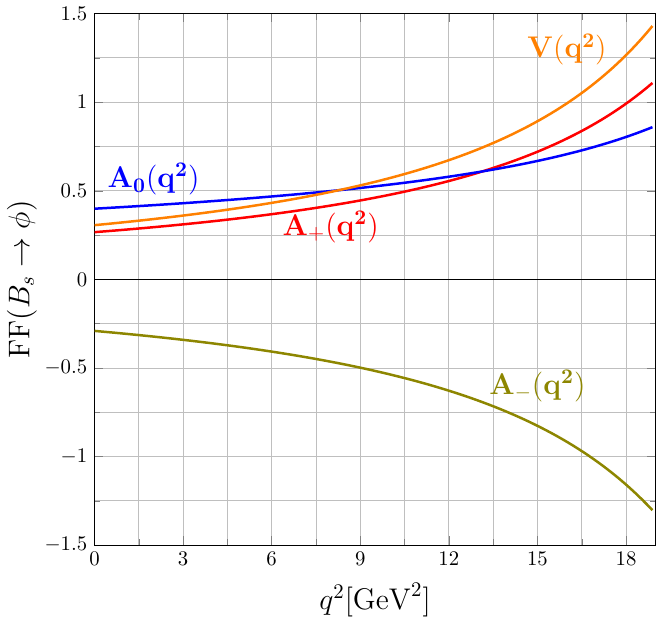}\quad{}\includegraphics[width=0.45\textwidth]{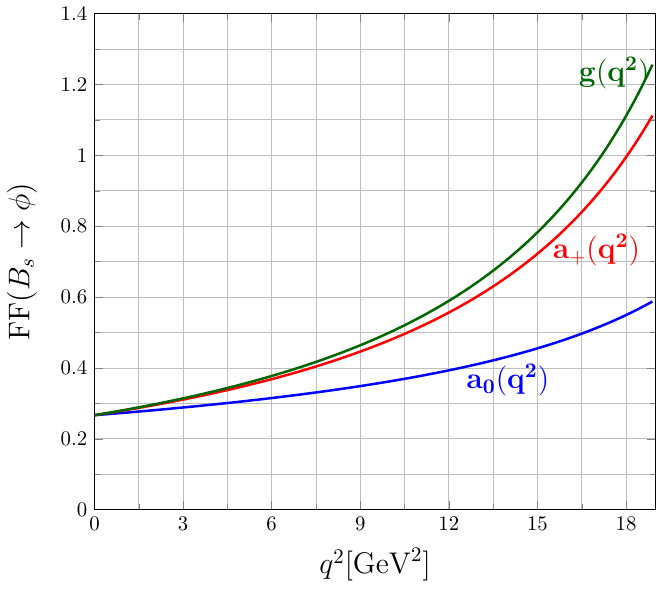}
\par\end{centering}
\caption{Form factors of the $B_{s}\to\phi$ transition.}
\label{Fig:FFs_Phi}
\end{figure}

The behavior of form factors is shown in Figs \ref{Fig:FFs_D} and
\ref{Fig:FFs_Phi}.

\subsection{Decay widths}

Once the decay constants and form factors are evaluated, branching
fractions can be determined: they represent observable quantities.
We compare our results to values published by the Particle Data Group
(PDG) \cite{ParticleDataGroup:2024cfk}.
\begin{center}
\begin{table}[H]
\begin{centering}
\begin{tabular}{lll}
 & CCQM & PDG\tabularnewline
\hline 
\noalign{\vskip0.2cm}
$\mathcal{B}(B_{s}^{0}\to D_{s}^{-}\,\pi^{+})$ & $5.107\times10^{-3}$ & $\left(2.98\pm0.14\right)\times10^{-3}$\tabularnewline
\noalign{\vskip0.2cm}
$\mathcal{B}(B_{s}^{0}\to D_{s}^{-}\,\rho^{+})$ & $13.49\times10^{-3}$ & $\left(6.8\pm1.4\right)\times10^{-3}$\tabularnewline
\noalign{\vskip0.2cm}
$\mathcal{B}(B_{s}^{0}\to D_{s}^{-}\,D^{+})$ & $4.89\times10^{-4}$ & $\left(3.1\pm0.5\right)\times10^{-4}$\tabularnewline
\noalign{\vskip0.2cm}
$\mathcal{B}(B_{s}^{0}\to K^{-}\,\pi^{+})$ & $6.06\times10^{-6}$ & $\left(5.9\pm0.7\right)\times10^{-6}$\tabularnewline
\noalign{\vskip0.2cm}
$\mathcal{B}(B_{s}^{0}\to\phi\,\overline{D^{0}})$ & $8.16\times10^{-6}$ & $\left(23.0\pm2.5\right)\times10^{-6}$\tabularnewline
\noalign{\vskip0.2cm}
$\mathcal{B}(B_{s}^{0}\to\phi\,\eta_{c})$ & $12.89\times10^{-4}$ & $\left(5.0\pm0.9\right)\times10^{-4}$\tabularnewline
\noalign{\vskip0.2cm}
$\mathcal{B}(B_{s}^{0}\to\phi\,J/\psi)$ & $1.697\times10^{-3}$ & $\left(1.03\pm0.04\right)\times10^{-3}$\tabularnewline
\noalign{\vskip0.2cm}
$\mathcal{B}(B_{s}^{0}\to K^{-}\,D^{+})$ & $9.376\times10^{-7}$ & \qquad{}-\tabularnewline
\noalign{\vskip0.2cm}
$\mathcal{B}(B_{s}^{0}\to K^{-}\,D^{*+})$ & $9.930\times10^{-7}$ & \qquad{}-\tabularnewline
\noalign{\vskip0.2cm}
$\mathcal{B}(B_{s}^{0}\to\phi\,D^{0})$ & $1.362\times10^{-6}$ & \qquad{}-\tabularnewline
\noalign{\vskip0.2cm}
$\mathcal{B}(B_{s}^{0}\to\phi\,\pi)$ & $1.555\times10^{-8}$ & \qquad{}-\tabularnewline
\noalign{\vskip0.2cm}
$\mathcal{B}(B_{s}^{0}\to D_{s}^{-}\mu^{+}\nu)$ & $2.697\times10^{-2}$ & $\left(2.29\pm0.21\right)\times10^{-2}$\tabularnewline
\noalign{\vskip0.2cm}
$\mathcal{B}(B_{s}^{0}\to K^{-}\mu^{+}\nu)$ & $1.196\times10^{-4}$ & $\left(1.06\pm0.09\right)\times10^{-4}$\tabularnewline
\noalign{\vskip0.2cm}
$\mathcal{B}(B_{s}^{0}\to\phi^{0}e^{+}e^{-})$ & $1.132\times10^{-6}$ & \qquad{}-\tabularnewline
\noalign{\vskip0.2cm}
\end{tabular}
\par\end{centering}
\end{table}
\par\end{center}

As mentioned before, we crosscheck the $B_{s}^{0}\to\phi\,J/\psi$
number above with our previous result from \cite{Dubnicka:2016nyy}
where $\mathcal{B}(B_{s}^{0}\to\phi\,J/\psi)=1.6\times10^{-3}$. We
see that our model gives consistent results over time, yet, some small
shifts can be caused by modification of the CCQM parameters when new
global fits are done to determine them. The decay $B_{s}^{0}\to\phi^{0}e^{+}e^{-}$
has no value for the total branching fraction, but numbers are know
\cite{LHCb:2024rto} for three individual $q^{2}$ intervals:
\begin{center}
\begin{table}[H]
\begin{centering}
\begin{tabular}{ccc}
\multicolumn{2}{c}{$10^{8}\times\frac{\mathcal{B}(B_{s}^{0}\to\phi^{0}e^{+}e^{-})}{dq^{2}}$} & $q^{2}$ interval (GeV)\tabularnewline
\hline 
\noalign{\vskip0.1cm}
CCQM & LHCb & \tabularnewline
\noalign{\vskip0.1cm}
$6.12$ & $13.8\pm3.1$ & $0.1<q^{2}<1.1$\tabularnewline
\noalign{\vskip0.1cm}
$3.30$ & $2.6\pm0.6$ & $0.1<q^{2}<6.0$\tabularnewline
\noalign{\vskip0.1cm}
$4.88$ & $3.9\pm1.2$ & $15<q^{2}<19$\tabularnewline
\noalign{\vskip0.1cm}
\end{tabular}
\par\end{centering}
\end{table}
\par\end{center}

\section{Discussion, summary}

The results give us a complex picture and one sees that for several
decays the description of the data is bad. Before the quantitative
comparison let us mention that the model has an intrinsic error and
errors on its input parameters. It is difficult to provide a well
justified estimate on how these impact the CCQM results: we attribute
our predictions a fixed 25\% uncertainty as a rough estimate. The
errors were analyzed in details for different decays in \cite{Issadykov:2025gkf}
(section IV) and we consider the findings there as a justification
for our choice, we do not perform a thorough analysis here. One may
notice that the uncertainties established in \cite{Issadykov:2025gkf}
are larger than the ones previously used (see e.g references 53 and
54 of \cite{Issadykov:2025gkf}).

The most precise results we get are represented by the semileptonic
decays: here the description is fine. Even disregarding the model
error the theoretical predictions are within the $2\sigma$ range
of the experimental uncertainties.

A satisfactory agreement with the measured values is observed for
$B_{s}^{0}\to D_{s}^{-}\,D^{+}$ and $B_{s}^{0}\to K^{-}\,\pi^{+}$,
it is within acceptable error range when both, model and data errors
are considered. These decays share the same diagrams but differ in
form factors. The numbers are quite off for $B_{s}^{0}\to D_{s}^{-}\,\pi^{+}$
and $B_{s}^{0}\to D_{s}^{-}\,\rho^{+}$, approximately by a factor
of 2. Relying on the theoretical uncertainty of 25\% and adding errors
in quadrature we get
\begin{align*}
B_{s}^{0}\to D_{s}^{-}\,\pi^{+}: & \left|\mathcal{B_{\text{CCQM}}-B_{\text{Ex.}}}\right|=(2.127\pm1.28)\times10^{-3},\\
B_{s}^{0}\to D_{s}^{-}\,\rho^{+}: & \left|\mathcal{B_{\text{CCQM}}-B_{\text{Ex.}}}\right|=(6.69\pm3.65)\times10^{-3}.
\end{align*}
Clearly, our model error estimate is dominant and important when drawing
a conclusion. If we trust it, we are at the limit of what can be accepted,
with the deviations being close to $2\sigma$, but not exceeding it.
At this point a comparison with other authors becomes interesting.
Most of them evaluate the charge-conjugated reaction.
\begin{center}
\begin{table}[H]
\centering{}%
\begin{tabular}{lll}
Decay & Branching fraction$\times10^{3}$ & Source\tabularnewline
\hline 
\noalign{\vskip0.1cm}
$\overline{B}_{s}^{0}\to D_{s}^{+}\,\pi^{-}$ & $2.98\pm0.14$ & Data (PDG)\tabularnewline
\noalign{\vskip0.1cm}
 & $2.8\:\pm\:?$ & \cite{Deandrea:1993ma}\tabularnewline
\noalign{\vskip0.1cm}
 & $4.39_{-1.19}^{+1.36}$ & \cite{Huber:2016xod}\tabularnewline
\noalign{\vskip0.1cm}
 & $4.42\pm0.21$ & \cite{Bordone:2020gao}\tabularnewline
\noalign{\vskip0.1cm}
 & $4.61_{-0.39}^{+0.23}$ & \cite{Cai:2021mlt}\tabularnewline
\noalign{\vskip0.1cm}
 & $2.15_{-1.20}^{+2.14}$ & \cite{Piscopo:2023opf}\tabularnewline
\hline 
\noalign{\vskip0.1cm}
$B_{s}^{0}\to D_{s}^{+}\,\rho^{-}$ & $6.8\pm1.4$ & Data (PDG)\tabularnewline
\noalign{\vskip0.1cm}
 & $7.5\:\pm\:?$ & \cite{Deandrea:1993ma}\tabularnewline
\noalign{\vskip0.1cm}
 & $11.30_{-3.11}^{+3.56}$ & \cite{Huber:2016xod}\tabularnewline
\noalign{\vskip0.1cm}
\end{tabular}
\end{table}
\par\end{center}

In general an overshooting of the experimental data is observed, which
we confirm. The two works which give numbers near to experimental
values have important errors: the number from \cite{Piscopo:2023opf}
may be compatible with $\mathcal{B}\sim4\times10^{-3}$ and the publication
\cite{Deandrea:1993ma} does not provide uncertainties for the values
we cite (Table 6b there). Yet, it provides errors for different decays
(see e.g. Table 6a) and they are very important, reaching often 50-100\%,
even more. We therefore presume that the value $2.8\times10^{-3}$
has also a very important error. When various models with different
form factor derivation point in the same direction one may get the
suspicion that the model itself cannot explain the discrepancies and
that the decays are not well understood, some ``physics'' is missing.
The authors of \cite{Bordone:2020gao} even entitle it as a ``puzzle''.
It might be related to factorization breaking (see discussion in \cite{Bordone:2020gao}
and \cite{Piscopo:2023opf}), new physics \cite{Cai:2021mlt} or some
other effect. 

From the three remaining decays that have experimental numbers, only
the $B_{s}^{0}\to\phi\,J/\psi$ is in rough agreement with the data
\[
B_{s}^{0}\to\phi\,J/\psi:\left|\mathcal{B_{\text{CCQM}}-B_{\text{Ex.}}}\right|=(0.67\pm0.43)\times10^{-3}.
\]
Decays $B_{s}^{0}\to\phi\,\overline{D^{0}}$ and $B_{s}^{0}\to\phi\,\eta_{c}$
have significant deviations from measurements, and in opposite directions.
Yet, they rely on the same form factor and share the color-suppressed
diagram, complemented by a penguin in the $\eta_{c}$ case. Again,
comparison with other theoretical evaluations and their factorization
approaches provides some insights (we do not distinguish between charge-conjugated
modes)
\begin{center}
\begin{table}[H]
\begin{centering}
\begin{tabular}{lll}
$\mathcal{B}(B_{s}^{0}\to\phi\,\overline{D^{0}})\times10^{6}$ & Factorization & Source\tabularnewline
\hline 
\noalign{\vskip0.1cm}
$23.0\pm2.5$ &  & Data (PDG)\tabularnewline
\noalign{\vskip0.1cm}
$31\pm7$ & Fact.-assisted topological-amplitude & \cite{Zhou:2015jba}\tabularnewline
\noalign{\vskip0.1cm}
$0.873\pm0.01$ & Na\"{i}ve & \cite{Talebtash:2016tym}\tabularnewline
\noalign{\vskip0.1cm}
$4.03_{-0.02}^{+0.01}$ & QCD factorization (QCDF) & \cite{Talebtash:2016tym}\tabularnewline
\noalign{\vskip0.1cm}
$5.47_{-0.91}^{+1.01}$ & Final-state interactions (FSI) & \cite{Talebtash:2016tym}\tabularnewline
\noalign{\vskip0.1cm}
$18.4_{-1.7}^{+1.8}$ & QCDF+FSI & \cite{Talebtash:2016tym}\tabularnewline
\noalign{\vskip0.1cm}
$19.7\pm4.5$ & Vertex + hard scatt. corr. (VHSC) & \cite{Bakhshi:2024hds}\tabularnewline
\noalign{\vskip0.1cm}
\end{tabular}
\par\end{centering}
\end{table}
\par\end{center}

One sees that the authors of \cite{Talebtash:2016tym} tested various
factorization techniques and got results which differ a lot. They
reach the agreement with the data only when QCDF and FSI approaches
are combined. The authors of \cite{Bakhshi:2024hds} tested the scale
dependence of the results for two methods (na\"{i}ve, VHSC) and have
shown an important scale dependence of the former, which is much less
pronounced in the VHSC case. Including our result, we cannot get a
fully coherent picture of the cause of existing variations but the
conclusion that the na\"{i}ve factorization we use is inappropriate
seems plausible. Besides questioning the theoretical evaluations,
it is also appropriate to point to the experimental situation: the
LHCb number has changed significantly, at the limit of consistency
\cite{LHCb:2018toh,LHCb:2023fqx}
\[
\mathcal{B}_{2018}^{\text{LHCb}}=\left(30.0\pm4.1\right)\times10^{-6},\qquad\mathcal{B}_{2023}^{\text{LHCb}}=\left(23.0\pm2.5\right)\times10^{-6}
\]
Thus an additional experimental confirmation of the number is desirable
before final conclusions are drawn.

The existing estimates for $B_{s}^{0}\to\phi\,\eta_{c}$ stand
\begin{center}
\begin{table}[H]
\begin{centering}
\begin{tabular}{ll}
$\mathcal{B}(B_{s}^{0}\to\phi\,\eta_{c})\times10^{4}$ & Source\tabularnewline
\hline 
\noalign{\vskip0.1cm}
$5.0\pm0.9$ & Data (PDG)\tabularnewline
\noalign{\vskip0.1cm}
$2.795\pm1.652$ & \cite{Ke:2017wni}\tabularnewline
\noalign{\vskip0.1cm}
$5.63_{-1.38}^{+1.86}$ & \cite{Xiao:2019mpm}\tabularnewline
\noalign{\vskip0.1cm}
\end{tabular}
\par\end{centering}
\end{table}
\par\end{center}

For this decay the published results are quite consistent with the
experiment, but small in number. It is therefore difficult to identify
the relevant differences that influence the result. The two displayed
works use different factorization approaches (na\"{i}ve vs. perturbative
QCD) but also different ways for hadronic transition amplitudes.

At last, we want to mention the work \cite{Ivanov:2023wir}, where
the CCQM is used to describe nonleptonic decays of baryons. The authors
take into the account long‐distance interactions through pole diagrams
which include intermediate baryon resonances. They conclude that the
pole diagrams are important to get consistency with measured data
and this suggests that description of nonleptonic decays without these
effects is not accurate enough.

In summary, we have in our work covered a wide area of $B_{s}$ decay
processes with various levels of agreement with data. We believe that
a large part of inconsistencies can be attributed to the na\"{i}ve
factorization breaking. The $B_{s}$ decays are often seen as complementary
to the $B_{d}$ ones. One can then follow the discussion regarding
the factorization for $B_{d}\to DX$ , which sometimes yields contradictory
conclusions, e.g. \cite{Neubert:2001sj} and \cite{Xing:2001nj}.
Nevertheless, the processes $B_{s}^{0}\to D_{s}^{-}\,\pi^{+}$ and
$B_{s}^{0}\to D_{s}^{-}\,\rho^{+}$ are among the cleanest non-leptonic
decays with no penguin or color-suppressed contributions. Yet, systematic
discrepancies are observed across several models, confirmed also by
us. This is an interesting observation that deserves focus and additional
analysis.

\bibliographystyle{ieeetr}
\bibliography{Bs_decays}

\end{document}